\documentclass{aa}  

\usepackage{graphicx} 
\usepackage{subfigure}
\usepackage{txfonts}
\usepackage{amsmath}
\usepackage{appendix}
\usepackage[hyperfootnotes=false]{hyperref}

\begin{document}

   \title{Capture and Stability of Resonant Planet Pairs in Turbulent Disk}

   \author{Linghong Lin\inst{1} \and Beibei Liu\inst{1 \star} \and Fei Dai\inst{2} \and Bin Liu\inst{1} \and Jiwei Xie\inst{3,4} \and Man Hoi Lee\inst{5,6,7} \and Haifeng Yang\inst{1} \and Shangfei Liu\inst{7,8}\and Ping Chen\inst{1}
          }

   \institute{Institute for Astronomy, School of Physics, Zhejiang University, Hangzhou 310027,  China\\
              \email{bbliu@zju.edu.cn}
              \and
              Institute for Astronomy, University of Hawai`i, 2680 Woodlawn Drive, Honolulu, HI 96822, USA
              \and
              School of Astronomy and Space Science, Nanjing University, Nanjing 210023, China
              \and
              Key Laboratory of Modern Astronomy and Astrophysics in Ministry of Education, Nanjing University, Nanjing 210093, China
              \and 
              Department of Earth and Planetary Sciences, The University of Hong Kong, Pokfulam Road, Hong Kong, China
              \and
              Department of Physics, The University of Hong Kong, Pokfulam Road, Hong Kong, China
              \and
              Hong Kong Institute for Astronomy and Astrophysics
              \and 
              School of Physics and Astronomy, Sun Yat-sen University, Zhuhai 519082, China
              \and 
              CSST Science Center for the Guangdong-Hong Kong-Macau Great Bay Area, Sun Yat-sen University, Zhuhai 519082, China
             }

   \date{}
 
  \abstract
   {We present a theoretical framework for the resonance capture and stability  of two-planet systems in turbulent disks. By incorporating stochastic forcing (parameterized by $\kappa$) alongside laminar angular momentum and eccentricity damping timescales ($\tau_{\rm m}, \tau_{e}$), we derive an analytical criterion for the general $j:j-1$ mean motion resonances, and validate it through N-body simulations. The outcome is mapped in $\kappa$-$\tau_{\rm m}/\tau_{e}$ parameter space, revealing two distinct regimes: resonance trapping and turbulence-induced disruption---which occurs either directly cross or via temporary capture followed by escape through turbulent diffusion. Crucially, our analysis identifies turbulence as a universal destabilizer. It amplifies the intrinsic overstability mechanism: 
   In laminar disks, escape requires $\tau_{\rm m}/\tau_{e}$ to drop below a critical limit due to excessive eccentricity excitation. We demonstrate that turbulent diffusion lowers this limit, demanding stronger damping (larger $\tau_{\rm m}/\tau_{e}$) for stability. Thus, greater turbulence promotes escape, and sufficiently strong diffusion precludes resonance retention irrespective of eccentricity damping.}


   \maketitle

\section{Introduction}

Type I migration for low-mass planets in gaseous protoplanetary disks is well established \citep{Goldreich&Tremaine1980, Ward+1997, Tanaka+2002, Kley&Nelson2012, Baruteau2013, Paardeekoper2023}. Slow convergent migration naturally drives planet pairs into mean-motion resonances (MMRs) \citep{Lee&Peale2002, Ogihara2013, Wang2014, Batygin2015, Huang2023, Batygin2023, Wong2024, Wang2024, Lin2025}. This paradigm is supported by \citet{Dai2024}, who found that observed young planetary systems are preferentially found in near-resonant configurations. However, observations of mature planetary systems reveal a broadly smooth period ratio distribution, exhibiting only minor deviations (excesses or deficits) near exact MMRs \citep{Fabrycky2014, Berger2020, Wang2021,Weiss2023}. This discrepancy points to the need for physical mechanisms that can disrupt initially resonant architectures \citep{Baruteau2013, Izidoro2017, Liu2017, Wuyq2024, Li2025}.

One promising mechanism is resonant overstability, a process where the resonant equilibrium point becomes unstable, causing small perturbations to grow exponentially. Originally developed in the context of Saturnian satellites \citep{Meyer2008}, this concept was applied by  \citet{Goldreich2014} to explain the scarcity of resonant exoplanets. Subsequent theoretical studies have extended this framework to the unrestricted three-body problem \citep{Deck&Batygin2015, Lin2025} and to second-order mean motion resonances \citep{Xu&Lai2017}. Furthermore, this mechanism has been investigated for specific planetary systems \citep{Nesvorny2022, Hu2025, Wang2026,Entrican2026, Batygin2026, Batygin2026B} and validated in hydrodynamic simulations \citep{Hands2018, Ataiee2021, Afkanpour2024}.

However, most previous studies assume a laminar disk, which represents an idealized simplification. Realistic protoplanetary disks are expected to be turbulent, driven by various mechanisms such as the magnetorotational instability (MRI; \citet{Balbus1991,Balbus1998,Beckwith2011, Simon2012, Rea2024}), gravitational instability (GI; \citet{Rice2003, Deng2017}), and the vertical shear instability (VSI; \citet{Nelson2013,Richard2016, Flock2017, Manger2020}). These instabilities generate transonic turbulent eddies, producing stochastic forcing through turbulence-induced density fluctuations \citep{Nelson2005, Wu2024, Kubli2025}. Such forcing introduces random perturbations that can significantly alter migration rates and destabilize resonant capture \citep{Zhou2002, Adams2008,rein2009, Okuzumi&Ormel2013, Batygin&Adams2017, Izidoro2017, Chen2025}.

To address the interplay between deterministic migration and stochastic perturbations, this study builds upon the framework established in \cite{Lin2025} (hereafter Paper I). In Paper I, we derived analytical criteria for resonant capture and stability strictly within the laminar regime. Here, we extend that theoretical model to turbulent environments by incorporating stochastic forcing terms into the equations of motion. By bridging the laminar and turbulent regimes, we aim to quantify how disk turbulence modifies the stability thresholds for resonant planetary pairs.

The paper is organized as follows. Section \ref{sec:model} presents our model, including prescriptions for both smooth and stochastic migration, and details the numerical setup. Section \ref{sec:result} derives an analytical criterion for turbulence-induced resonance disruption and presents $N$-body simulations that validate these predictions. Finally, Sections \ref{sec:discuss} and \ref{sec:conclusion} discuss our results in the context of previous studies and summarize the main conclusions.

\section{Model}
In this section, we describe the prescriptions of smooth migration torque and stochastic torque in Sections \ref{sec:mig_torque} and~\ref{sec:sto_torque}, respectively.  
The numerical setup is presented in Section \ref{sec:num_setup}. \label{sec:model}
\subsection{Smooth migration torque} 
\label{sec:mig_torque}
Following Paper I, the dissipative effect of the planet-disk interaction in a $2$D laminar disk can be parameterized as \citep{Teyssandier2014, Ataiee&Kely2021, Pichierri2023}
\begin{subequations} \label{disk_ef}
\begin{eqnarray}
      \frac{1}{L}\frac{dL}{dt} & = & -\frac{1}{\tau_{\rm m}}, \\
   \frac{1}{a}\frac{da}{dt} & = & -\frac{1}{\tau_{\rm a}} =   -\frac{2}{\tau_{\rm m}} - \frac{2e^{2}}{1-e^{2}}\frac{1}{\tau_{e}},
    \label{eq:da_dt2} \\
      \frac{1}{e}\frac{de}{dt} & = & -\frac{1}{\tau_{e}} \label{de_dt},
\end{eqnarray} 
\end{subequations}
where the orbital elements $L$, $a$, $e$, denote the angular momentum, semi-major axis and eccentricity, $\tau_{\rm m}$, $\tau_{\rm a}$ and $\tau_{\rm e}$ represent the damping timescales for orbital angular momentum, semi-major axis and eccentricity, respectively. While $\tau_{\rm a}$ is a derived quantity, our formalism strictly treats $\tau_{\rm m}$ and $\tau_{\rm e}$ as the two independent input parameters.  The accelerations for angular momentum and eccentricity damping are given by \citep{Papaloizou2000, Cresswell2006, Cresswell2008} 
    \begin{eqnarray}
       {\textbf{a}_{\rm m}}  =  - \frac{\textbf{v}}{\tau_{\rm m}},  \ \ 
       {\textbf{a}_{\rm e}} =  -2 \frac{(\textbf{v} \cdot \textbf{r}) \cdot \textbf{r}}{r^2 \tau_{e}},
    \end{eqnarray}
where $\textbf{v}$ and $\textbf{r}$ are the planet's velocity and radius vectors.  

\subsection{Stochastic torque}
\label{sec:sto_torque}
In addition to the nominal gravitational potential of the laminar disk where gas follows a smooth distribution and regular sub-Keplerian motion, turbulence induces random fluctuations in gas density and velocity. The resulting turbulent gravitational potential exerts additional stochastic forces on embedded planets. Following \citet{rein&choksi2022}, we model this effect as an extra stochastic acceleration term in the planet's equation of motion.

The strength of this turbulent forcing is quantified by the dimensionless parameter  $\kappa {=} \sqrt{\langle \textbf{a}^2_{\rm sto} \rangle} /\textbf{a}_{\star}$, defined as the ratio of the root-mean-square (RMS) stochastic acceleration to the gravitational acceleration from the central star. This can be written as 
\begin{equation}
    \sigma_{\textbf{\rm sto}} =
    \kappa\frac{GM_{\star}}{a^{2}}, \label{eq:kap_def}
\end{equation}
where the stationary variance  $\sqrt{\langle \textbf{a}_{\rm sto}^{2} \rangle}{ =} \sigma_{\rm sto}$.
A higher $\kappa$ corresponds to a more turbulent disk and thus stronger stochastic perturbations on the planet. Physically, $\kappa$ can be related to the disk's turbulent viscosity parameter $\alpha$ via established diffusion formulations \citep{rein2012, Okuzumi&Ormel2013, Batygin&Adams2017}, which will be further discussed in Section~\ref{sec:discuss}.

The stochastic acceleration is modeled as a smoothly varying process that changes the strength and direction of $\textbf{a}_{\rm sto}$ over time.
\begin{equation}
\frac{d \textbf{a}_{\rm{sto}}}{dt}
= -\frac{\textbf{a}_{\rm{sto}}}{\tau_{\rm c}}
+ \sqrt{\frac{2\,\sigma_{\textbf{\rm sto}}^{2}}{\tau_{\rm c}}}\,\xi(t), 
\label{a_sto}
\end{equation}
where $\tau_{\rm c}$ is the correlation timescale of turbulent fluctuations. In our simulations, we adopt $\tau_{\rm c} = 2\pi/n$ (i.e., one orbital period) \citep{Adams2008,rein2009, oishi2007}, where $n = \sqrt{GM_{\star}/a^3}$ is the mean motion of the planet. $\xi(t)$ is a Gaussian white noise process with zero mean and unit variance satisfying $\langle \xi(t)\,\xi(t') \rangle = \delta(t - t')$.
The first term on the right side of Eq.~\eqref{a_sto} describes the gradual damping of the turbulent acceleration over the timescale $\tau_{\rm c}$, while the second term introduces new random perturbations. Taken together, these terms ensure that $\textbf{a}_{\rm sto}$ evolves smoothly in time and retains only short-term memory over $\tau_{\rm c}$, reflecting the finite lifetime of turbulent eddies in the disk. In other words, this formulation captures the gradual, time-evolving nature of turbulent forcing rather than representing it as a series of instantaneous random kicks.

The cumulative effect of these stochastic forces leads to a random walk in the planet's orbital elements. Specifically, the tangential component of the stochastic acceleration exerts a fluctuating torque on the planet, driving a continuous diffusion in its angular momentum $L$. Utilizing the relation $n \propto L^{-3}$ (which implies $dn/dL = -3n/L$) and the scaling of the stochastic amplitude $\sigma_{\rm sto} = \kappa n^2 a$, the variance of the mean motion fluctuations grows linearly with time as:
\begin{equation}
   \langle (\Delta n)^2 \rangle = \left( \frac{dn}{dL} \right)^2 \langle (\Delta L)^2 \rangle = 9\kappa^2 n^4 \tau_{\rm c} t.
    \label{eq:diffusion_n}
\end{equation}
This formulation allows us to quantify the intensity of orbital diffusion, serving as a critical basis for the stability analysis in Section \ref{sec:tur_crit}.

\subsection{Numerical setup}
\label{sec:num_setup}
We conduct numerical simulations using the open-source N-body code \texttt{REBOUND} with the \texttt{WHfast} integrator \citep{Rein&Liu2012}. The orbits of the planets are assumed to be coplanar. The planet-disk interaction terms accounting for angular-momentum and eccentricity damping, as well as stochastic acceleration, were implemented through the \texttt{REBOUNDx} \citep{Tamayo2020, rein&choksi2022}.

In this study, we specifically focus on the two-planet pair system at the 2:1 mean-motion resonance. However, since the analytical framework developed in this work follows the formulation of Paper I, the results can be naturally extended to general resonance configurations. Initially, the inner planet is placed at $1\ \rm au$, and the outer planet starts at $1.7\ \rm au$, slightly exterior to the exact 2:1 commensurability. Both planets are initialized on circular orbits with random phase angles. Same as Paper I, in order to mimic the effect of convergent migration, only the outer planet undergoes inward migration on a timescale of $\tau_{\rm m}$, while the eccentricities of both planets are damped on a timescale of $\tau_{e}$. Each simulation is integrated up to $t{=}\tau_{\rm m}$ unless otherwise stated.

\begin{figure}
    \centering
    \subfigure{
        \includegraphics[width=0.45\textwidth]{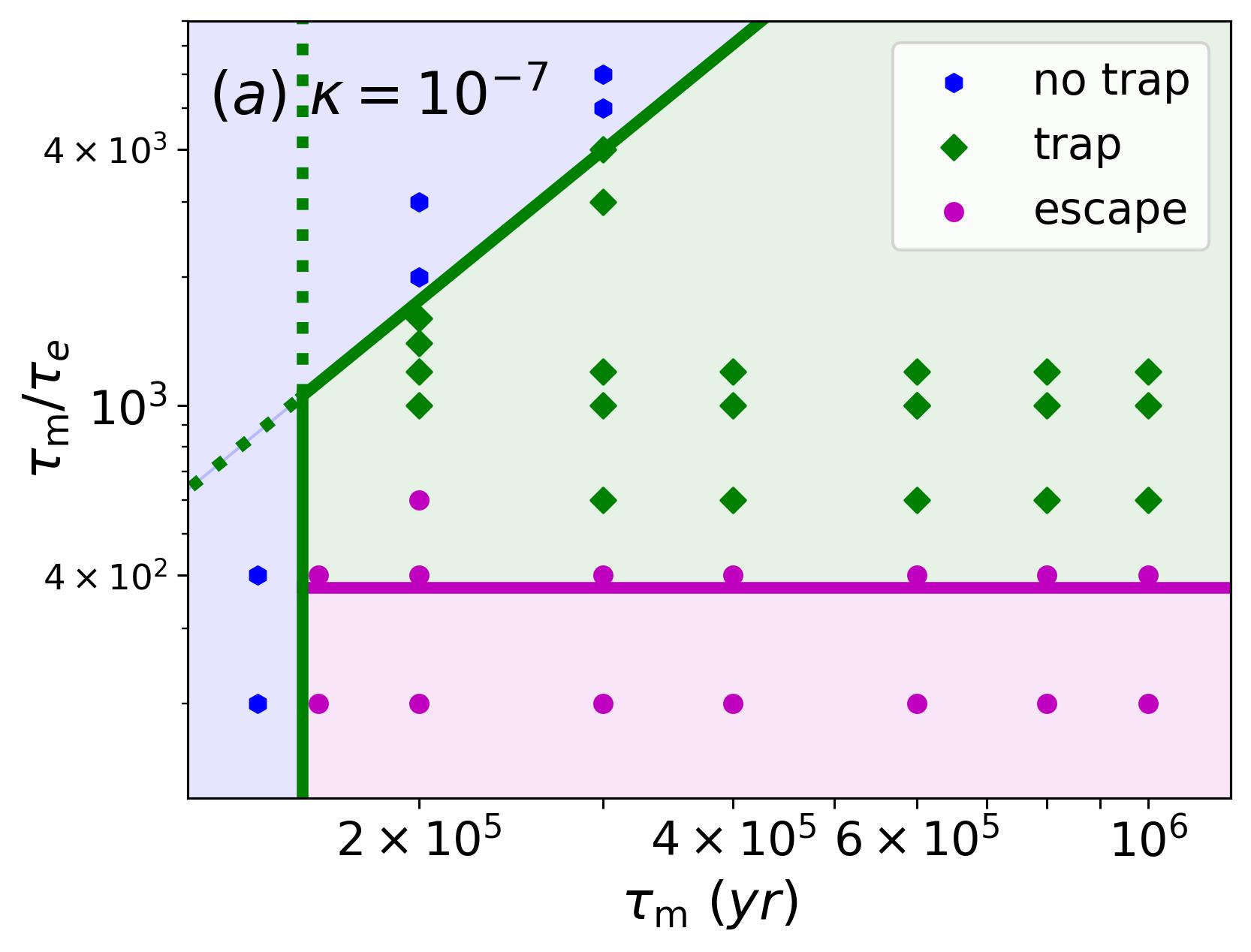}
        \label{fig:k=1e-7}
    }\hspace{0.01\textwidth}
    \subfigure{
        \includegraphics[width=0.45\textwidth]{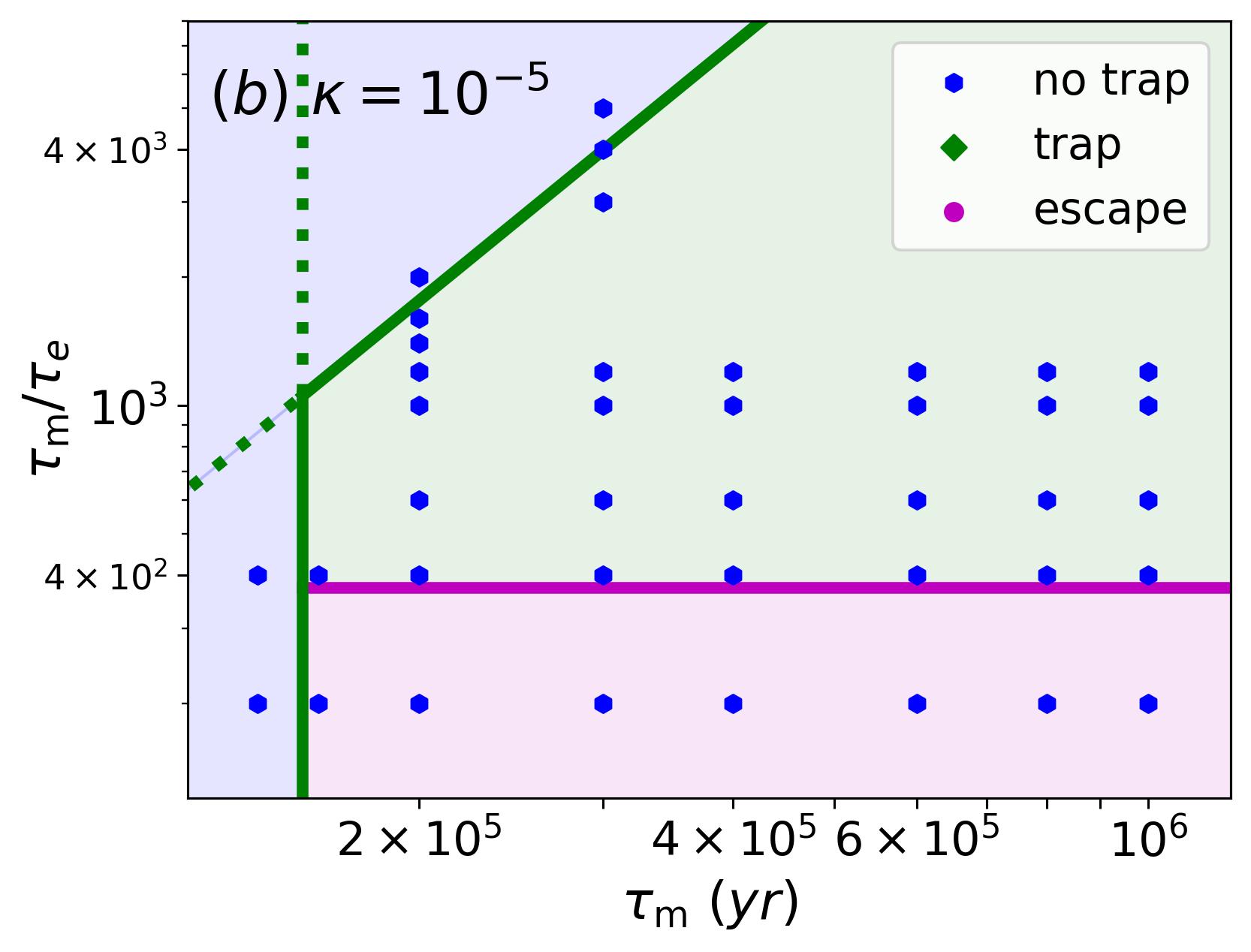}
        \label{fig:k=1e-5}
    }
    \caption{Simulations of the $2$:$1$ MMR capture and stability for a planet-pair at three different disk turbulent strengths: (a) $\kappa {=} 10^{-7}$ and (b) $10^{-5}$. The planet masses are $m_{\rm i}{=}1\ M_{\oplus}$, $m_{\rm o}{=}10\ M_{\oplus}$. Solid lines represent our analytical criteria for a laminar disk, while colored dots correspond to numerical results. While weak turbulence ($\kappa = 10^{-7}$) preserves the laminar-like behavior, strong turbulence ($\kappa = 10^{-5}$) introduces dominant stochastic forcing that completely disrupts resonance stability.}
    \label{fig:diff_kap}
\end{figure}

\section{Results}
\label{sec:result}
In this section, we present our numerical and analytical  results and explore how disk turbulence influences resonance dynamics. We consider two planets of masses $m_{\rm i}$ and $m_{\rm o}$ orbiting a central star of $M_{\star}$ on coplanar orbits. The normalized masses and the mass ratio between the inner and outer planets are $\mu_{\rm i,o} {= }m_{\rm i,o} / M_{\rm \star}$ and $q{=}m_{\rm i}/m_{\rm o}$,  respectively. For notation, the subscripts ‘i’ and ‘o’ refer to the corresponding quantities of the inner and outer planets.

In Section~\ref{sec:qual_analy}, we perform demonstration simulations to illustrate how stochastic forcing modifies the system's resonance behavior. In Section~\ref{sec:tur_crit}, we derive the new analytical criteria for the transition between resonance trapping and disruption with the consideration of turbulent diffusion. Finally, we compare these analytical expressions with $N$-body simulations in Section~\ref{sec:num_setup}.

\subsection{Qualitative Analysis}
\label{sec:qual_analy}
Here we qualitatively show how the disk turbulence influences the system’s resonant dynamics. Paper I constructed an analytical framework describing resonant dynamics in laminar disks. As a first step, we examine how the different stochastic forcing strengths affect the above analysis. We adopt two representative values in our simulations: $\kappa = 10^{-7}$ and $10^{-5}$. We focus on a low-mass inner planet configuration with $m_{\rm i} = 1 \ M_{\oplus}$, $m_{\rm o} = 10 \ M_{\oplus}$.

Figure~\ref{fig:diff_kap} shows the numerical results with two $\kappa$ values. The colored dots represent the numerical outcomes, and for comparison the lines correspond to the analytical criteria derived from Paper I without taking into account disk turbulence. The shaded regions, partitioned by these analytical boundaries, represent the predicted capture outcomes (see Figure 6 in Paper I). Here we collectively refer to stable and overstable resonant trap as 'trap', since the turbulent diffusion effectively erases the distinction between these two regimes (also see Figure \ref{fig:kap_phase}). 

In the weak-turbulence case ($\kappa {= }10^{-7}$, Figure~\ref{fig:diff_kap}a), the simulation results align closely with the analytical predictions of the laminar-disk framework. However, when $\kappa$ increases to $10^{-5}$, the stochastic forcing becomes strong enough to make all systems vulnerable to resonance disruption. Their behavior deviates significantly from the laminar-disk analysis (Figure~\ref{fig:diff_kap}b).

Next, we examine how the stability of the systems depends on $\kappa$. We focus on the case with $\tau_{\rm m} = 6 \times 10^{5}$\,yr and $\tau_{\rm m}/\tau_{e} = 10^{3}$ in Figure \ref{fig:diff_kap}, and analyze the evolution of the planets' period ratio as a function of $\kappa$ in Figure \ref{fig:kap_period}. When $\kappa$ is low (e.g., $10^{-7}$, orange line), the planets remain in resonance throughout the simulation, exhibiting behavior nearly identical to the case without turbulence (blue line). As $\kappa$ increases to $10^{-6}$ (green line), the planets are initially trapped in resonance but escape after $5\times 10^{5}$\,yr. Consequently, the duration of resonance capture decreases with increasing $\kappa$. For a sufficiently high $\kappa$ of $5 \times 10^{-6}$ (brown line), the planets cross the resonance directly without capture. This trend clearly indicates that the system's stability progressively decreases with increasing $\kappa$. We provide further theoretical analysis to elucidate the effects of disk turbulence in the next subsection.

\begin{figure}
    \centering
    \includegraphics[width=1\linewidth]{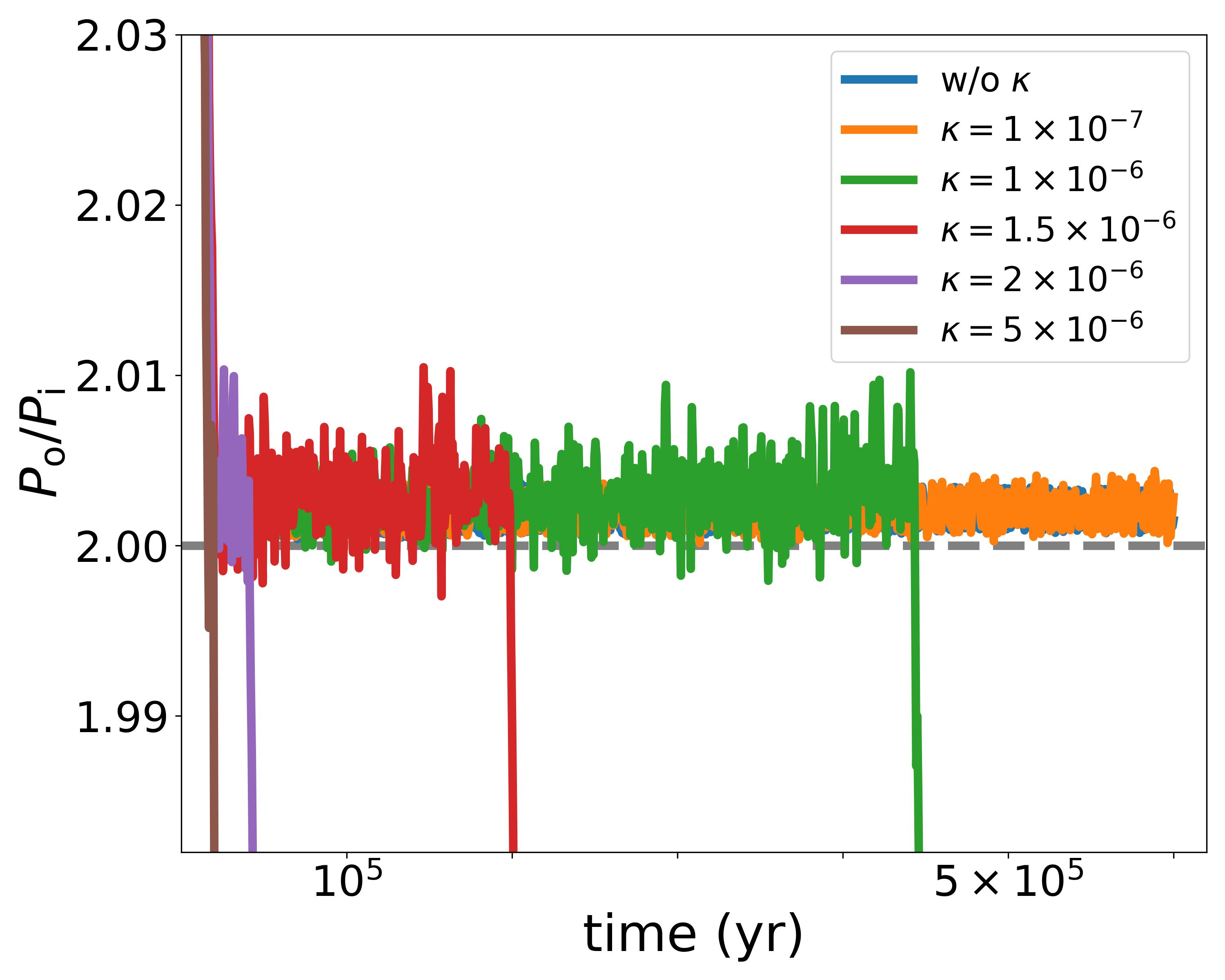}
    \caption{Time evolution of the period ratio ($P_{\rm o}/P_{\rm i}$) for a planetary pair converging towards the $2$:$1$ MMR with $\tau_{\rm m} = 6 \times 10^{5}$\,yr and $\tau_{\rm m}/\tau_{e} = 10^{3}$. The colored lines represent different turbulence strengths $\kappa$, and the grey dotted line represents the nominal resonant location. As the turbulence strength increases, the resonance is disrupted more rapidly.
\label{fig:kap_period}}
    \label{fig:kap_period}
\end{figure}

\subsection{Analytical Criteria for Resonance Stability}
\label{sec:tur_crit}
Here we derive an analytical criterion for turbulence-induced resonance disruption as follows.

Turbulence introduces stochastic fluctuations to the orbital elements, driving a diffusive evolution that can disrupt the resonant configuration. To quantify this effect, we define the diffusivity $D_{A}$, which measures the growth rate of the variance of a parameter $A$, as \citep{Adams2008,rein2009, Okuzumi&Ormel2013}
\begin{equation}
    D_{A} \equiv \frac{\langle(\Delta A)^2\rangle}{\Delta t}.
\end{equation}
The diffusivity of a derived quantity $f(\{x_i\})$ is governed by the general error propagation formula for correlated variables \citep{Bevington2003}:
\begin{equation}
    D_{f} = \sum_i \left(\frac{\partial f}{\partial x_i}\right)^2 D_{x_i} + \sum_{i \neq j} \left(\frac{\partial f}{\partial x_i}\right)\left(\frac{\partial f}{\partial x_j}\right) \rho_{ij}\sqrt{D_{x_i}D_{x_j}},
    \label{eq:general_propagation}
\end{equation}
where $\rho_{ij}$ is the correlation coefficient between variables $x_i$ and $x_j$. Physically, a non-zero $\rho_{ij}$ arises in realistic disks when the characteristic eddy scale exceeds the planet separation, inducing spatially correlated stochastic perturbations. In this work, we assume that the stochastic perturbations acting on the two planets are spatially uncorrelated ($\rho_{\rm i,o} {=} 0$). Consequently, the cross-correlation terms vanish.

For a planetary pair near a first-order $j$:$j-1$ MMR, we employ the proximity parameter $x \equiv j/(j-1) - n_{\rm i}/n_{\rm o}$ to quantify the deviation from exact MMR. Substituting the diffusivity of the mean motion $D_n$ (Eq. \ref{eq:diffusion_n}), the diffusivity of $x$ is given by
\begin{equation}
\begin{split}
  D_{x} &= \left(\frac{\partial x}{\partial n_{\rm i}}\right)^2 D_{n_{\rm i}} + \left(\frac{\partial x}{\partial n_{\rm o}}\right)^2 D_{n_{\rm o}} \\
  &= 9\kappa^2 \left(\frac{n_{\rm i}}{n_{\rm o}}\right)^2 (n_{\rm i}^2 \tau_{\rm c,i} + n_{\rm o}^2\tau_{\rm c,o}).
\end{split}
\label{eq:diff_x}
\end{equation}
Here $D_x$ represents the variance growth rate of the resonance offset under a free-random-walk approximation.

Intuitively, the resonance is maintained only if the stochastic variation does not exceed the width of the resonance. The characteristic resonance width is defined as (see Appendix A for derivation):
\begin{equation}
    x_{\rm res} = 3\frac{n_{\rm i}}{n_{\rm o}} \frac{\tau_{\rm lib}}{\tau_{\rm slow, mig}},
    \label{eq:Delta_x}
\end{equation}
where $\tau_{\rm lib}$ is the libration timescale, and $\tau_{\rm slow, mig}$ represents the critical convergent migration timescale for resonance capture in a laminar disk (see Eqs. 8 and A6 in Paper I).

Turbulent forcing drives stochastic fluctuations in the resonance offset
\(x\). However, because the evolution of \(x\) is regulated by resonant
dynamics and eccentricity damping, it cannot be strictly treated as a free
random walk with indefinitely growing variance. Instead, the diffusive
spread is expected to approach a quasi-stationary level. We therefore write
the variance of \(x\) in the resonant state more generally as
\begin{equation}
\sigma_x^2 =
\int_0^{t_{\rm sat}}
D_{\rm eff}(\kappa,\tau_e,\tau_{\rm lib},\ldots;t)\,{\rm d}t ,
\label{eq:sigma_x_general}
\end{equation}
where \(D_{\rm eff}\) is the effective diffusivity after filtering by the resonant dynamics and eccentricity damping, and \(t_{\rm sat}\) is the characteristic saturation time.

In the analytical estimate below, we do not attempt to specify the full functional form of \(D_{\rm eff}\). Instead, we approximate the stochastic spread by assuming locally linear variance growth with the free-random-walk rate \(D_x\) over the resonant response time \(\tau_{\rm lib}\). This gives a leading-order estimate of the stochastic spread within resonance
\begin{equation}
    \sigma_x^{2} \simeq D_x\tau_{\rm lib} .
    \label{eq:sigma_x}
\end{equation}

In contrast to \citet{Batygin&Adams2017}, who assumed an unbounded random walk over the migration timescale \(\tau_{\rm m}\) (\(\sigma_x^2 \approx D_x \tau_{\rm m}\))---akin to an undamped stochastically forced pendulum \citep{Mallick&Marcq2004}---we argue for a constrained physical picture. Instead, the stochastic spread is estimated over the local resonant response time $\tau_{\rm lib}$.
\\
Despite this saturation, resonance destabilization remains an irreversible process. Over long timescales, the system inevitably samples the extreme tails of this Gaussian distribution. These rare, large excursions create deviations from exact resonance that significantly exceed the standard deviation $\sigma_x$. To analytically capture these separatrix-crossing events, we introduce a safety factor $\epsilon$ and adopt the criterion that resonance is disrupted when the effective diffusive excursion exceeds the resonance width:
\begin{equation}
    \epsilon \sigma_x > x_{\rm res}.
\end{equation}
Substituting Eqs. \ref{eq:sigma_x} and \ref{eq:Delta_x} into this condition, we obtain the analytic criterion for turbulence-induced disruption as
\begin{equation}
    \kappa \gtrsim \frac{1}{\epsilon} \frac{\sqrt{\tau_{\rm lib}}}{\tau_{\rm slow,mig}} \frac{1}{\sqrt{\left(n_{\rm i}^2 \tau_{c,i} + n_{\rm o}^2\tau_{c,o}\right)}}.
    \label{eq:kappa_crit}
\end{equation}
Our numerical results indicate that $\epsilon {\approx} 3$ provides stability over a timescale $\tau_{\rm m}$, consistent with the standard $3\sigma$ for the Gaussian distribution. This corresponds to a resonance width that encloses roughly $99.7\%$ of stochastic variations, reducing escape probability to a low level. While stochastic diffusion mathematically causes eventual escape over an infinite timescale ($\Delta t {\to} \infty$), this criterion provides a practical measure of long-term stability over the
timescales considered here.

This practical criterion should be interpreted as a leading-order estimate. It assumes locally linear variance growth over the resonant response time, \(\sigma_x^2 \simeq D_x\tau_{\rm lib}\), and does not explicitly include eccentricity damping in the effective diffusion. The possible effect of \(\tau_e\) on resonance disruption is examined in Appendix~B.

The condition above describes direct resonance disruption by stochastic excursions that exceed the resonance width. Turbulence can also destabilize the system indirectly in the moderate-turbulence regime, where the system remains nominally bounded ($\epsilon \sigma_x {<} x_{\rm res}$), stochastic forcing can continuously amplify the overstability until it ultimately drives the system to escape resonance. In the laminar case, stability depends on the balance between eccentricity excitation (driven by convergent migration $\tau_{\rm m}$) and eccentricity damping ($\tau_{e}$). The ratio $\tau_{\rm m}/\tau_{e}$ serves as the key parameter: as this ratio decreases, eccentricity excitation dominates and the system transitions from stable to overstable, and eventually escape (see Fig. 6 of paper I). Resonant escape occurs when $\tau_{\rm m}/\tau_{e}$ falls below a critical threshold derived in Paper I:
\begin{equation}
    \frac{\tau_{\rm m}}{\tau_{e}} < \left(\frac{\tau_{\rm m}}{\tau_{e}}\right)_{\rm esc} = \left(\frac{3}{\mu_o}\right)^{2/3} \frac{h(q)}{4} \left[ \frac{(j-1)^{2}+j^{2}q}{f_{\rm d,i}\alpha + f_{\rm d,o}^{'}q^{2}} \right]^{2/3},
    \label{eq:laminar_criterion}
\end{equation}
where 
\begin{equation}
    h(q) = \frac{1}{(1+q\sqrt{\alpha})} \frac{f_{\rm d,i}^{2}+f_{\rm d,o}^{'2}q^{2}\alpha}{j f_{\rm d,i}^{2}+(j-1)f_{\rm d,o}^{'2}q\sqrt{\alpha}}.
\end{equation}
Here, $\alpha {\equiv} a_{\rm i}/a_{\rm o}$ denotes the semi-major axis ratio (distinct from the viscosity parameter $\alpha_{\rm SS}$ used in Section \ref{sec:discuss}). The coefficients $f_{\rm d,i}$ and $f_{\rm d,o}$ represent the direct terms of the disturbing function; their detailed expressions are provided in Table 8.1 of \citet{Murray1999}. The modified outer coefficient is defined as $f_{\rm d,o}^{'} \equiv f_{\rm d,o} - 2\alpha \delta_{j,2}$. For instance, in the case of the $2$:$1$ MMR, these coefficients take the values $f_{\rm d,i} \approx -1.19$ and $f_{\rm d,o}^{'} \approx 0.43$.

\begin{figure}
    \centering
    \includegraphics[width=1.1\linewidth]{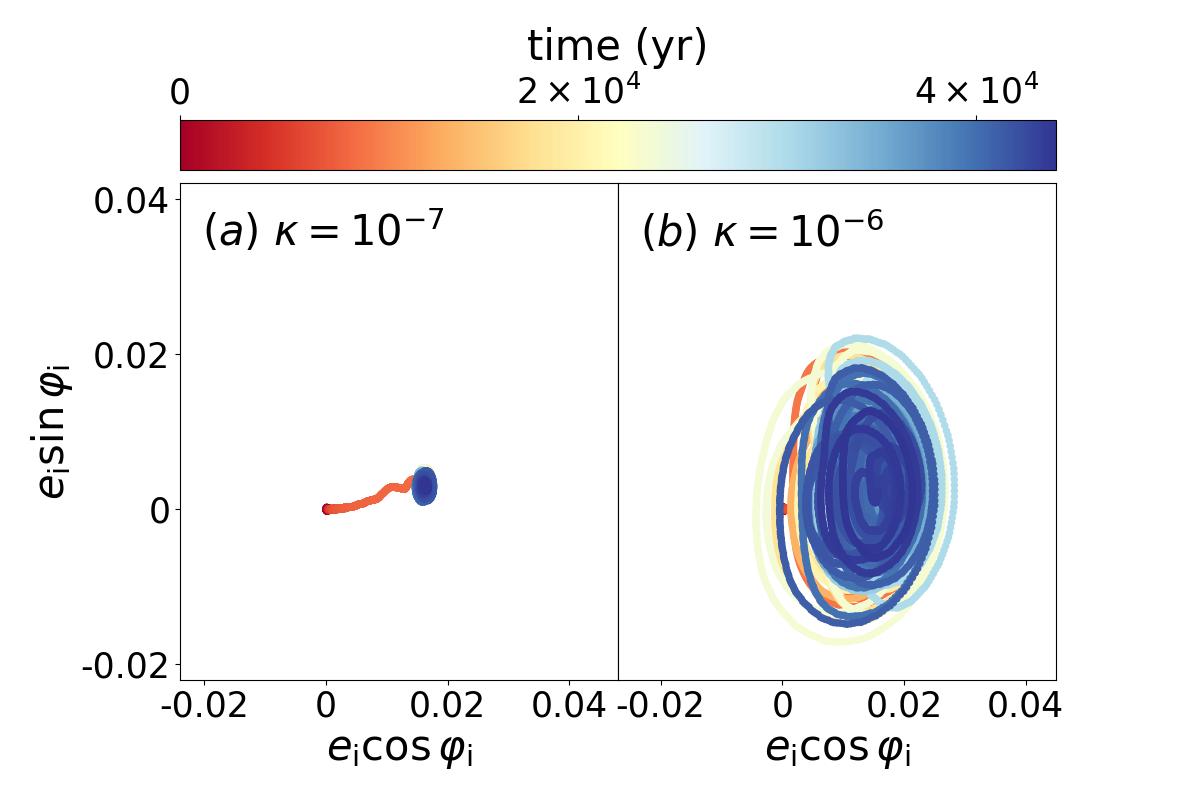}
    \caption{Evolution of a two-planet system in phase space at two different disk turbulent strengths: (a) $\kappa {=} 10^{-7}$ and (b) $\kappa {=} 10^{-6}$. The resonant angle is defined as $\varphi_{\rm i} = j\lambda_{\rm o} - (j-1)\lambda_{\rm i} - \varpi_{\rm i}$, where $\lambda$ is the mean longitudes of the planet, and $\varpi$ is the relevant longitude of pericenter. The planet mass and migration parameters are $m_{\rm i} {=} 1\ M_{\oplus}$, $m_{\rm o} {=} 10\ M_{\oplus}$, $\tau_{\rm m} {=} 6 \times 10^{5}$  yr and $\tau_{\rm m} / \tau_{e} {=} 10^{3}$. Turbulence induces orbital eccentricity diffusion around the equilibrium point, with stronger diffusion observed at a higher $\kappa$ value.}
    \label{fig:kap_phase}
\end{figure}

In a turbulent disk, stochastic diffusion effectively amplifies overstability. The underlying mechanism is illustrated in Figure \ref{fig:kap_phase}: under conditions of weak turbulence ($\kappa=10^{-7}$), the system remains localized near the equilibrium point. However, as turbulence increases ($\kappa=10^{-6}$), the trajectory explores an increasingly wider area of the phase space. The combined action of stochastic dispersion and resonant overstability drives the trajectory toward the separatrix, ultimately allowing the system to escape its captured state far more readily than predicted by standard laminar theory.

To quantify this effect, we introduce a function $f_{\rm tur} {=} \max[0, 1 - \epsilon \sigma_x/x_{\rm res}]$ into the laminar-limit formula, which yields a new disruption criterion as
\begin{equation}
  f_{\rm tur}  \frac{\tau_{\rm m}}{\tau_{e}} < \left(\frac{\tau_{\rm m}}{\tau_{e}}\right)_{\rm esc}.
    \label{eq:turbulent_criterion}
\end{equation}
When $\epsilon \sigma_{x}{\gtrsim}x_{\rm res}$, $f_{\rm tur}$ approaches zero and the above equation is always satisfied by any given $\tau_{\rm m}$ and $\tau_{e}$.  When $\epsilon \sigma_{\rm x}{\ll}x_{\rm res}$, $f_{\rm tur}$ approaches unity, returning back to the laminar disk circumstance. 

\begin{figure}
    \centering
    \includegraphics[width=1\linewidth]{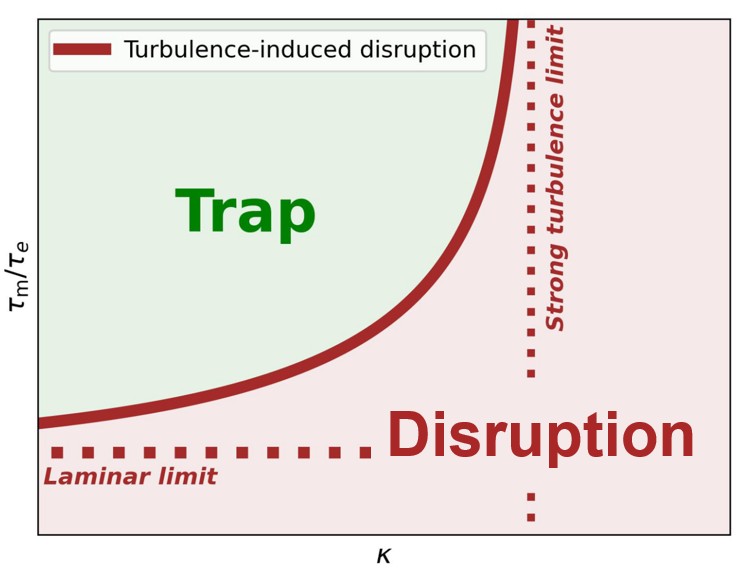}
    \caption{Analytical predictions for the convergent migration of a planet pair near the 2:1 MMR in a turbulent disk. The green region denotes stable resonance trapping, while the brown region corresponds to resonance disruption (resulting in either direct crossing or transient capture followed by escape). The solid brown line represents the turbulence-induced criterion (Eq. \ref{eq:final_kappa}) that separates these two regimes. The horizontal dotted line marks the laminar limit, and the vertical dotted line indicates the strong turbulence limit (Eq. \ref{eq:kappa_crit}). For a typical sub-Neptune system ($\sim 10$ $M_{\oplus}$), this strong turbulence limit corresponds to $\kappa_{\mathrm{crit}} \approx 10^{-6}$.} 
    \label{fig:illustration}
\end{figure}

Finally, by substituting  Eq. \eqref{eq:sigma_x} into Eq. \eqref{eq:turbulent_criterion} and solving for the turbulence strength $\kappa$, we derive the critical turbulent strength $\kappa$  for resonance disruption as
\begin{equation}
    \kappa \gtrsim \frac{1}{\epsilon} \frac{\sqrt{\tau_{\rm lib}}}{\tau_{\rm slow,mig}} \frac{1}{\sqrt{\left(n_{\rm i}^2 \tau_{c,i} + n_{\rm o}^2\tau_{c,o}\right)}} \left[ 1 - \frac{(\tau_{\rm m}/\tau_{e})_{\rm esc}}{\tau_{\rm m}/\tau_{e}} \right].
    \label{eq:final_kappa}
\end{equation}
This result can be directly compared to \citet{Batygin&Adams2017}, who derived a criterion consistent with our strong-turbulence limit (cf. our Eq. 12 with their Eq. 13). This comparison reveals a key distinction rooted in the diffusive timescale ($\tau_{\rm lib}$ vs $\tau_{\rm m}$), as anticipated from the discussion following Eq. \ref{eq:sigma_x}. More importantly, our formulation provides a unified criterion that bridges laminar and turbulent regimes, offering a continuous description across different disk conditions.

Figure \ref{fig:illustration} summarizes our generalized framework in the $\kappa$--$\tau_{\rm m}/\tau_{e}$ parameter space. The solid brown curve delineates the analytical boundary (Eq. \ref{eq:final_kappa}) that separates the resonance trapping regime (green region) from the resonance disruption regime (brown region).

In the former stable regime and when the disk is weakly turbulent ($\kappa {\to} 0$), the above boundary line asymptotically converges to the escape criterion in the laminar disk (horizontal dotted line). Here, turbulence is too weak to directly lead to resonance crossing. Overstability grows as $\kappa$ gradually increases. The boundary line then deviates upward, indicating that stronger eccentricity damping (larger $\tau_{\rm m}/\tau_{e}$) is required to avoid overstablity. 
As $\kappa$ further increases, the boundary curve terminates at a vertical asymptote (vertical dotted line), corresponding to the critical $\kappa$ derived in Eq. \eqref{eq:kappa_crit} for resonance crossing.  We note that the unstable regime can either refer to the directly resonance crossing (to the right of the vertical line) or temporarily capture but escape later (to the left). Due to the diffusive nature of turbulence, we do not distinguish these two types and call them together as turbulence-induced  disruption.  

\subsection{Comparison with simulations}
\label{sec:num_setup}
\begin{figure}
    \centering
    \includegraphics[width=0.99\linewidth]{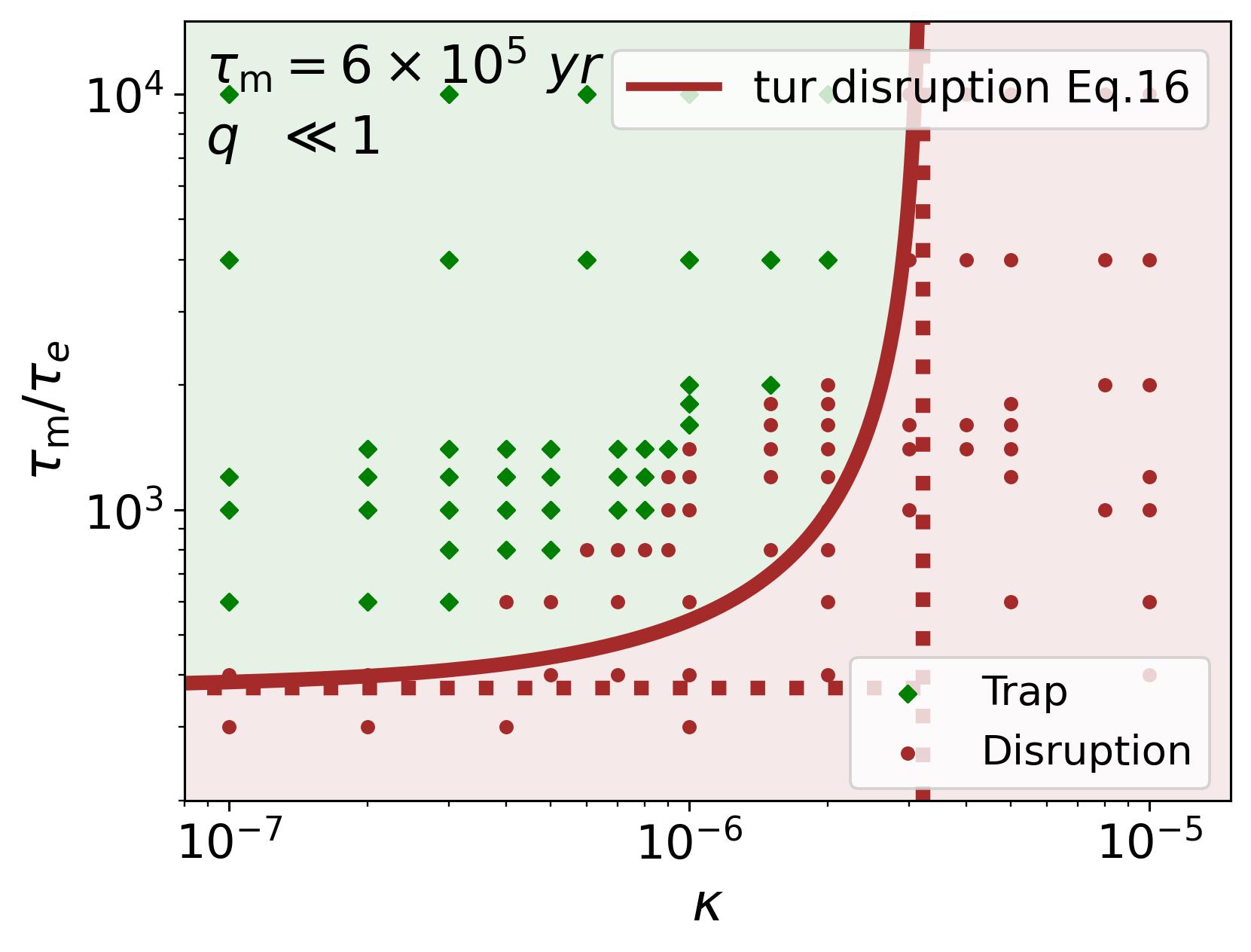}
    \caption{Simulations of 2:1 MMR capture and stability in a turbulent disk for a planet pair with a mass ratio $q=0.1$. The solid brown line represents the turbulence-induced disruption criterion (Eq. \ref{eq:final_kappa}). For comparison, the horizontal and vertical dotted lines mark the laminar limit (Eq. 24 in Paper I) and the strong turbulence limit (Eq. \ref{eq:kappa_crit}), respectively. The simulation outcomes show broad consistency with the analytical predictions }
    \label{fig:kap_crit}
\end{figure}

\begin{figure}
    \centering
    \includegraphics[width=0.99\linewidth]{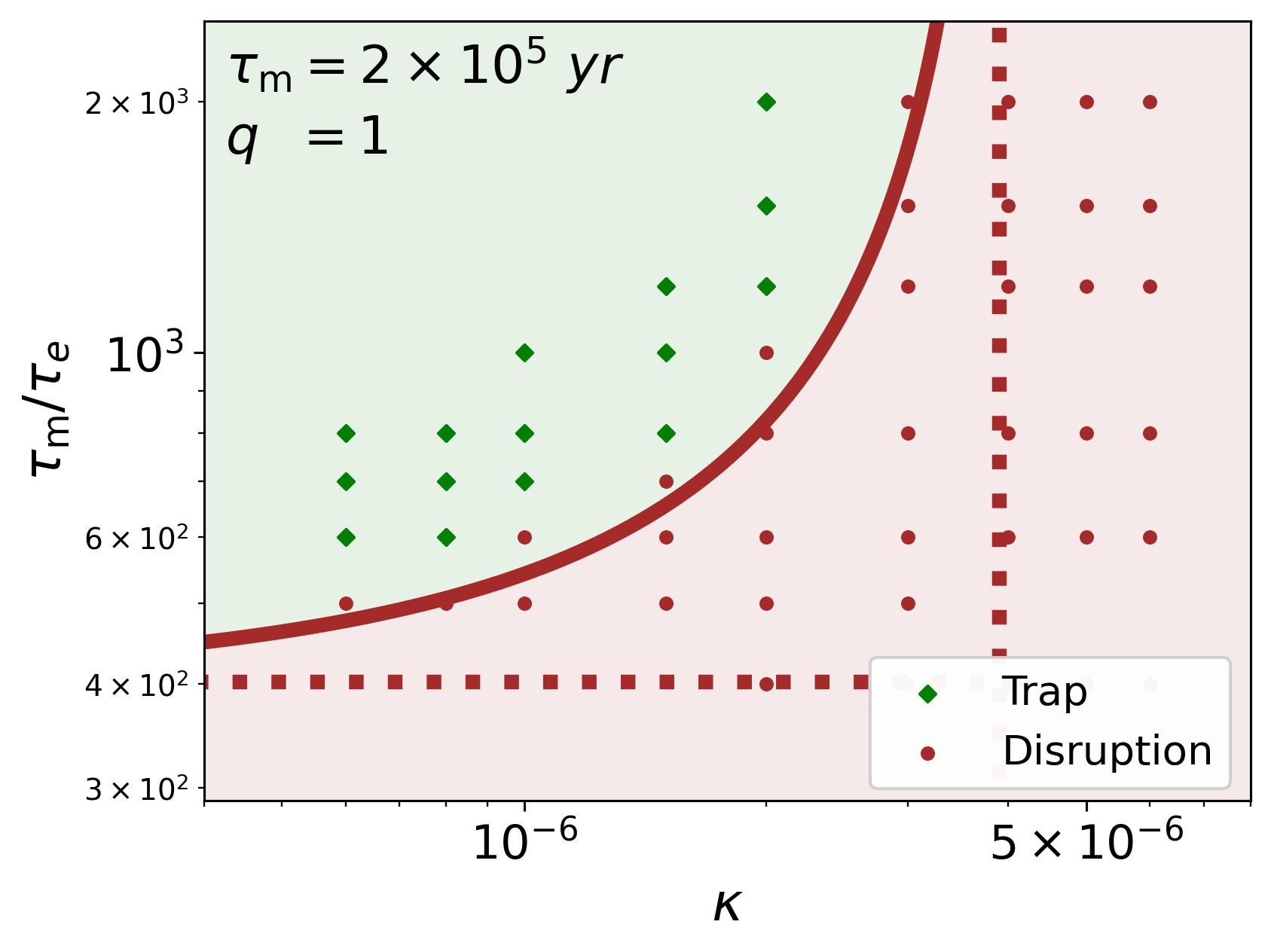}
    \caption{Same as Figure \ref{fig:kap_crit} but with inner to outer planet mass ratio $q=1$.}
    \label{fig:kap_crit_q_1}

    \vspace{1cm}

    \includegraphics[width=0.99\linewidth]{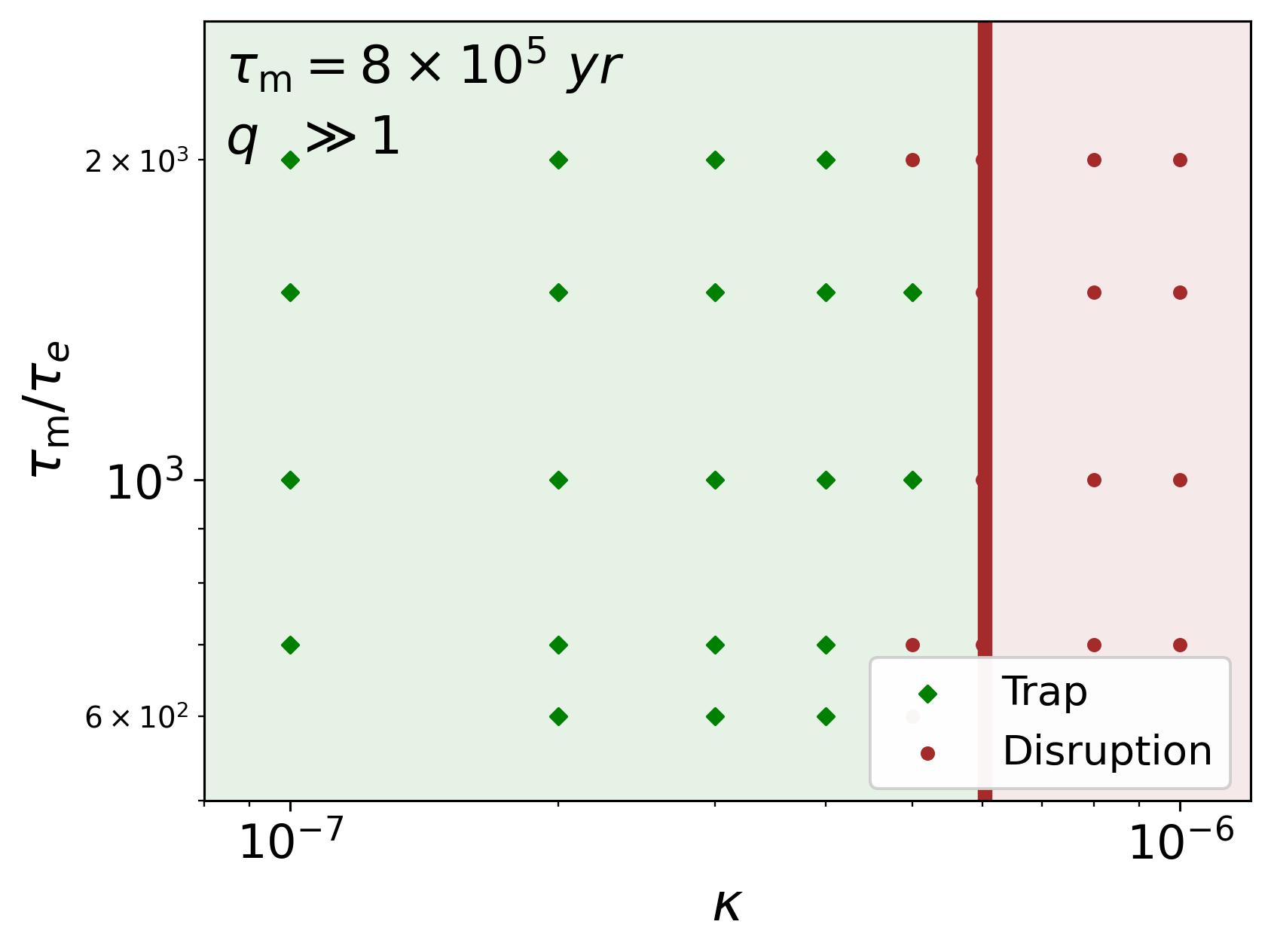}
    \caption{Same as Figure \ref{fig:kap_crit} but with inner to outer planet mass ratio $q=10$.}
    \label{fig:kap_crit_q_10}
\end{figure}

We perform numerical simulations to validate the derived criterion (Eq. \ref{eq:final_kappa}). The results are presented in Figures \ref{fig:kap_crit}, \ref{fig:kap_crit_q_1}, and \ref{fig:kap_crit_q_10}, covering three representative mass ratio regimes: small inner planet ($m_{\rm i} = 1 \, M_{\oplus}$, $m_{\rm o} = 10 \, M_{\oplus}$, $q = 0.1$, $\tau_{\rm m} = 6 \times 10^{5} \, {\rm yr}$), equal-mass planets ($m_{\rm i} = 10 \, M_{\oplus}$, $m_{\rm o} = 10 \, M_{\oplus}$, $q = 1$, $\tau_{\rm m} = 2 \times 10^{5} \, {\rm yr}$), and small outer planet ($m_{\rm i} = 10 \, M_{\oplus}$, $m_{\rm o} = 1 \, M_{\oplus}$, $q = 10$, $\tau_{\rm m} = 8 \times 10^{5} \, {\rm yr}$).

Overall, we find excellent agreement between the analytical predictions and the simulation outcomes. As illustrated in Figure~\ref{fig:kap_crit}, green dots denote systems that retain resonance at the end of the simulation, while brown dots indicate systems where resonance has been disrupted. Taking the small inner planet case ($q=0.1$) as a representative example, we observe that in the weak-turbulence limit ($\kappa \lesssim 10^{-7}$), stochastic perturbations are negligible, and the stability boundary asymptotically recovers the laminar-disk criterion derived in Paper I (horizontal dotted line). As $\kappa$ increases, intensified stochastic diffusion tends to destabilize the resonance. Consequently, stronger eccentricity damping (i.e., a larger $\tau_{\rm m}/\tau_{e}$) is required to counterbalance this effect and sustain capture. This mechanism offers a physical explanation for the phenomenon reported by \citet{Chen2025}, where turbulence was observed to expand the parameter space susceptible to overstability. Finally, when $\kappa \gtrsim 2 \times 10^{-6}$, the system crosses the critical strong-turbulence threshold (vertical dotted line, Eq.~\ref{eq:kappa_crit}), leading to direct resonance disruption. 
A minor discrepancy appears near $\kappa \approx 10^{-6}$, where the simulations exhibit unexpected disruption. This arises because our analytical model is based on a linear approximation, which breaks down when the system is pushed near the resonance separatrix. In this region, non-linear dynamics accelerate the escape, a detailed process not captured by our simplified diffusion model. Despite this, the analytical criterion remains a robust predictor of the overall stability trend.

The simulations also confirm the mass ratio dependence. Overstability exists for systems with $q {\le} 1$ (Figures \ref{fig:kap_crit} and \ref{fig:kap_crit_q_1}). Here, turbulence exacerbates the intrinsic overstability, thereby diminishing the system's overall stability and shifts the escape threshold to progressively larger $\tau_{\rm m}/\tau_{\rm e}$. In contrast, for systems with $q {\gg} 1$ (Figure \ref{fig:kap_crit_q_10}), overstability never occurs. The systems remain in permanent capture  even as turbulence reduces the effective resonance width. Consequently, this resonance capture is independent of the eccentricity damping. Disruption is driven exclusively by strong turbulence when $\kappa$ exceeds the critical threshold. Thus, the stable boundary is set by a vertical line (Eq. \ref{eq:kappa_crit}), distinct from the $q {\lesssim} 1$ regime.

Together, these results provide a clear understanding of how stochastic perturbations govern the long-term evolution and stability of resonances in turbulent protoplanetary disks.

\section{Discussion}
\label{sec:discuss}
In this section, we first establish a physical link between the stochastic forcing parameter $\kappa$ used in this work and the widely-adopted turbulent viscosity parameter $\alpha_{\rm SS}$ in Section~\ref{sec:alpha}. We then compare our results with previous studies in Section~\ref{sec:compare}. Finally, model limitations are discussed in Section~\ref{sec:caveat}.

\subsection{Connection to viscosity parameter $\alpha_{\rm SS}$}
\label{sec:alpha}
In this study, we adopt the classical prescription of \citet{Shakura1973}, where the turbulent viscosity is expressed as $\nu = \alpha_{\rm SS} H^{2} \Omega$, $H$ is the gas scale height and $\Omega$ is the Keplerian angular frequency. While $\alpha_{\rm SS}$ was originally formulated to describe angular momentum transport, the same underlying physics allows it to serve as an effective proxy for the amplitude of turbulent density fluctuations---the source of the stochastic gravitational forces that perturb planetary orbits \citep{Okuzumi2011}.

To relate $\alpha_{\rm SS}$ to $\kappa$, we adopt the framework of \citet{Okuzumi&Ormel2013}. In their model, which assumes ideal MHD turbulence, the diffusion coefficient for the semi-major axis can be expressed as 
$D_{a} \simeq \frac{\alpha_{\rm SS}}{2}\left(\frac{\Sigma a^{2}}{M_{\star}}\right)^{2} n^{2} a^{2}\tau_{\rm c}$ (their Eq.~50),
where $\Sigma$ is the gas surface density and $n$ is the mean motion. 
Converting this to our formulation via the relation $D_{n} = (9n^2/4a^2) D_{a}$ 
yields 

\begin{equation}
\alpha_{\rm SS} \simeq 10^{-4} \left(\frac{\kappa}{10^{-6}}\right)^{2} \left(\frac{M_{\star}}{M_{\odot}}\right)^{2} \left(\frac{\Sigma}{1700g\,cm^{-2}}\right)^{-2}\left(\frac{a}{1\ \rm au}\right)^{-4}.
\label{eq:alpha_kappa}
\end{equation}
To quantify this relationship, we adopt the Minimum Mass Solar Nebula (MMSN) profile \citep{Hayashi1981}, $\Sigma {=}1700 (a/1\ \rm au)^{-3/2}\,\mathrm{g\,cm^{-2}}$.

\begin{figure}
    \centering
    \subfigure{
        \includegraphics[width=0.5\textwidth]{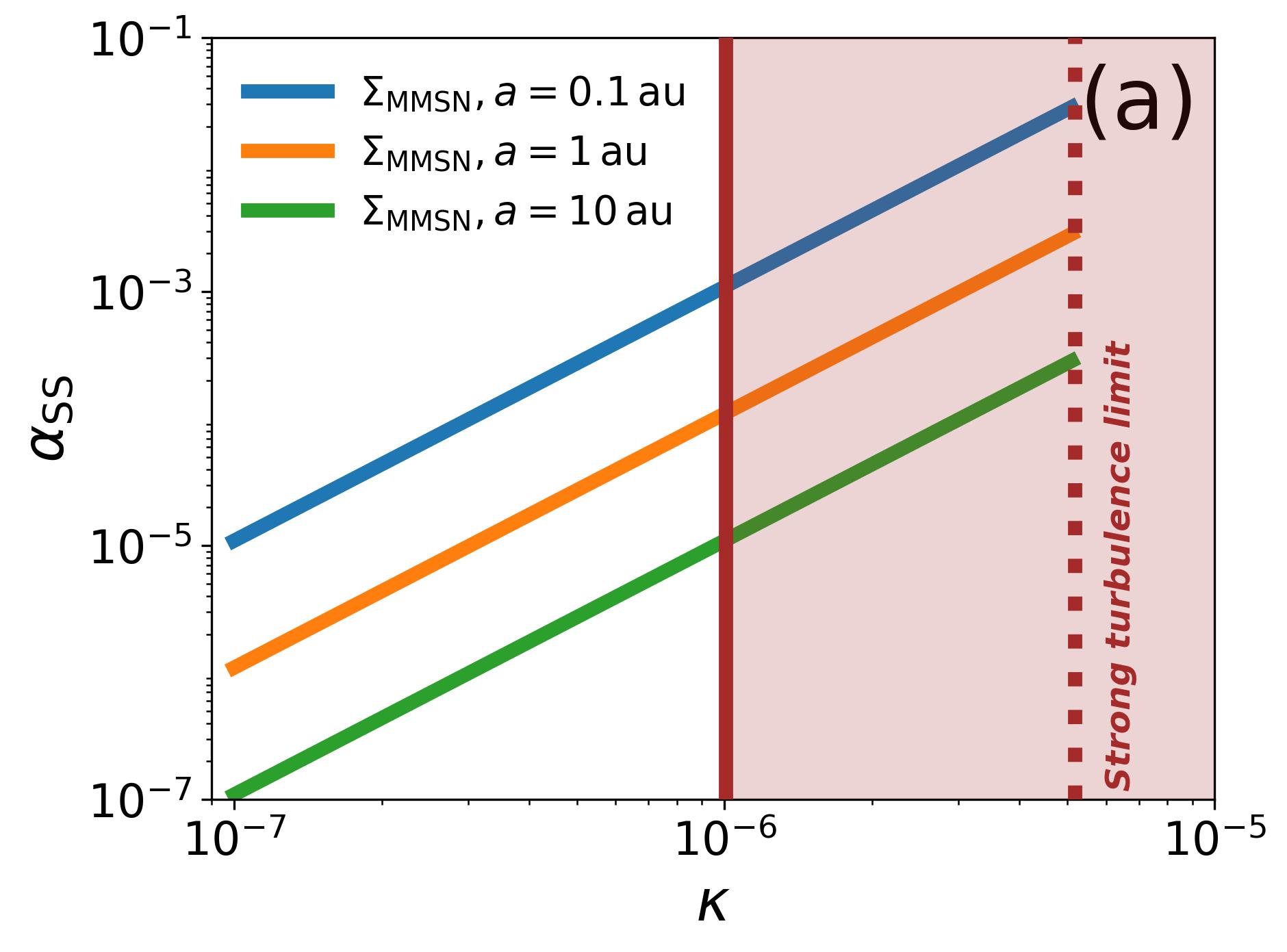}
        \label{fig:k_alpha}
    }\hspace{0.01\textwidth}
    \subfigure{
        \includegraphics[width=0.5\textwidth]{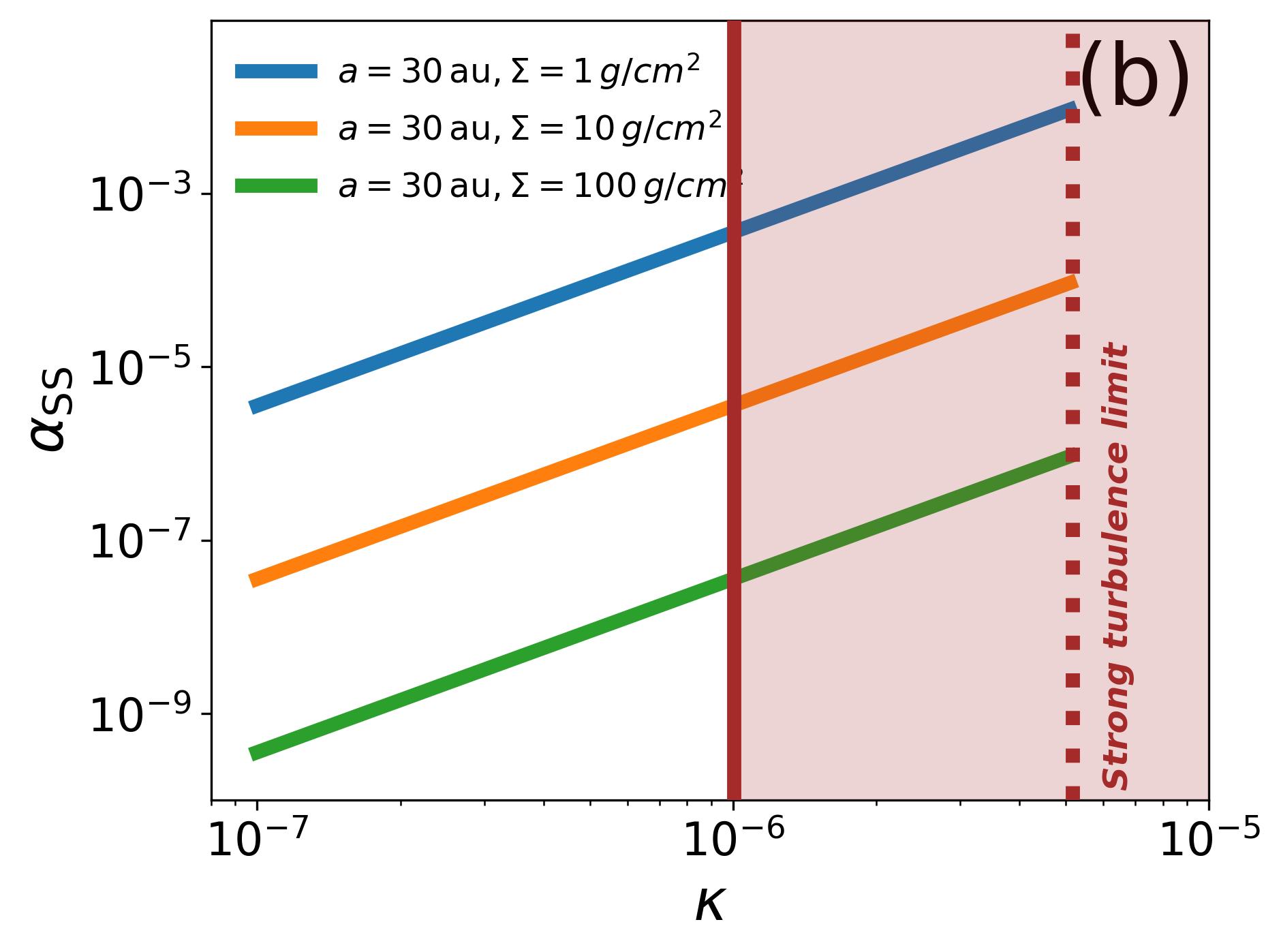}
        \label{fig:k_alpha_30au}
    }
    \caption{Correspondence between the dimensionless stochastic force amplitude $\kappa$ and the effective turbulent viscosity parameter $\alpha_{\rm SS}$.
    The results are calculated for a representative two-planet system with inner mass $m_{\rm i} = 1\,M_{\oplus}$ and outer mass $m_{\rm o} = 10\,M_{\oplus}$.
    (a) The relationship assuming a Minimum Mass Solar Nebula (MMSN) surface density profile with varying semi-major axis $a$.
    (b) The relationship at a fixed orbital distance of $a=30\,\mathrm{au}$ with varying gas surface densities $\Sigma$.
    The solid brown line depicts the critical turbulence strength required for resonance disruption for $\tau_{\rm m}/\tau_{e}=500$ (based on Eq.~\ref{eq:final_kappa}), while the dotted line marks the strong turbulence limit (Eq.~\ref{eq:kappa_crit}).}
\end{figure}

Figure~\ref{fig:k_alpha} illustrates the relationship between $\kappa$ and $\alpha_{\rm SS}$ for a representative two-planet system with inner mass $m_{\rm i} = 1\,M_{\oplus}$ and outer mass $m_{\rm o} = 10\,M_{\oplus}$. Our analytic criterion indicates that resonances become unstable at $\kappa \approx 10^{-6}$. In the context of the Minimum-Mass Solar Nebula (MMSN), this value of $\kappa$ corresponds to $\alpha_{\rm SS} \approx 10^{-3}$ at $0.1\,\mathrm{au}$, $\alpha_{\rm SS} \approx 10^{-4}$ at $1\,\mathrm{au}$, and $\alpha_{\rm SS} \approx 10^{-5}$ at $10\,\mathrm{au}$. These low-to-moderate $\alpha_{\rm SS}$ values are consistent with recent observations from young protoplanetary disks \citep{Pinte2016, Flaherty2017}, which suggests that turbulence-induced disruption can be relevant for resonance breaking in realistic disk environments. Crucially, this reveals a spatial dependence for MMR survival. As the critical $\alpha_{\rm SS}$ threshold drops by orders of magnitude with increasing distance from the star, sub-Neptune resonant architectures are far more likely to be preserved in the inner disk ($\sim 0.1\,\mathrm{au}$) than in the outer regions.

We also consider the case at $30$ au in Figure~\ref{fig:k_alpha_30au} for comparison with disk observation.  Across the observed surface density range of $\Sigma{\sim}1{-}100\,\mathrm{g\,cm^{-2}}$, the critical disruption threshold ($\kappa \approx 10^{-6}$) corresponds to extremely low viscosities, with $\alpha_{\rm SS} {\lesssim} 10^{-4}$.This indicates that resonances in the outer disk are especially susceptible to even modest levels of turbulence.

We emphasize that the viscous parameter here refers specifically to the efficiency of angular‑momentum transport. In the literature, $\alpha$ is also used to denote other effects, such as the mass diffusivity coefficient $\alpha_{\rm D}$ \citep{Youdin2007}.  We note that for MRI-driven turbulence---which is approximately isotropic---angular momentum transport is roughly equivalent to mass diffusion \citep{Johansen2006,Zhu2013}. Other driving mechanisms, however, may produce anisotropic turbulence, in which case $\alpha_{\rm SS}{ \neq} \alpha_{\rm D}$. Caution is therefore required to distinguish the viscous parameter that drives disk evolution from specific mixing parameters that govern gas and dust diffusivity \citep{Okuzumi2011, Stammler2022, Ying2026}. 

Furthermore, non-ideal MHD effects are expected to substantially suppress midplane turbulence, giving rise to dead zones. In such regions, angular momentum transport is dominated by disk winds, while the midplane remains relatively laminar  \citep{Bai2013, Bai2016, Gole2016, Yang2018, Lesur2023}. Although this decouples stochastic diffusion from large-scale accretion, Equation~\ref{eq:alpha_kappa} still provides an order-of-magnitude estimate of the forcing amplitude required to disrupt resonances.

\subsection{Comparison with Previous Studies}
\label{sec:compare}
To model the effects of turbulence on planet migration, early investigations typically parameterized potential fluctuations, $\Phi_{\rm turb}$, via an amplitude parameter $\gamma$ \citep{Laughlin2004, Ogihara2007, Baruteau2010, Izidoro2017, Huhn2021, Wu2024, Chen2025}. More recently, \citet{Soto&Zhu2025} analyzed global MHD simulations and introduced a diffusion coefficient, $C_{\rm D}$, to quantify the turbulence strength by directly measuring the variance of the instantaneous migration torque. In this work, we implement stochasticity within N-body simulations using a direct force model characterized by $\kappa$, defined as the ratio of the stochastic force magnitude to the stellar gravitational force \citep{rein&choksi2022}. Crucially, these different approaches—the potential-based $\gamma$, the torque-based $C_{\rm D}$, and our force-based $\kappa$—are fundamentally equivalent representations of the same underlying physical process and can be inter-converted via the disk's viscosity parameter $\alpha_{\rm SS}$ \citep{Adams2008, rein2009,Okuzumi&Ormel2013}.

Our findings broadly align with previous work. For instance, \citet{Izidoro2017} suggested that disk turbulence had minimal impact on breaking the resonance chain  based on their N-body simulations adopted $\alpha_{\rm SS}$ of $5.4 \times 10^{-3}$. Our criterion predicts a critical turbulence strength of $\kappa {\simeq} 2 \times 10^{-6}$ for super-Earths of $10 \ M_{\oplus}$.  This can be converted to an equivalent  $\alpha_{\rm SS}$ through Eq. \ref{eq:alpha_kappa} based on their disk model ($\Sigma {\sim} 3000\,{\rm g\,cm^{-2}}$ at 1 au and ${\sim} 2.5\times 10^{4}\,{\rm g\,cm^{-2}}$ at 0.1 au). At $r{=}0.1$ AU, the critical value derived from our criterion is $4 {\times} 10^{-2}$, much higher than the assumed $\alpha_{\rm SS}$ in their simulations, rendering the system insensitive to turbulent perturbations. However,  at $r{=}1$ au, the derived critical value is $3 \times 10^{-4}$, lower than the simulation value.  This indicates that resonance is more prone to disruption at larger orbital distances. This r-dependent resonant breaking feature is consistent with the results shown in their Figure 9.

Our analysis is also consistent with the recent hydrodynamic simulations of \citet{Chen2025}, who reported that turbulence expands the parameter space for overstability, causing planet pairs to migrate into tighter resonances. This agreement suggests that our framework provides a reasonable description of the essential physics behind these complex hydrodynamic outcomes.

Finally, in the specific case of the Kepler-223 system, \citet{Huhn2021} demonstrated through N-body simulations that the observed resonant chain configuration is reproduced most successfully assuming disk surface densities of a few MMSN and viscosity parameters $\alpha_{\rm SS}$ ranging from a few $10^{-3}$ up to $10^{-2}$. According to Eq. \ref{eq:alpha_kappa}, this upper limit of $\alpha_{\rm SS}$ corresponds to a turbulence strength of $\kappa \lesssim 10^{-6}$, which is, again in agreement with our analytical criterion.

\subsection{Model limitations}
\label{sec:caveat}
Here, we outline the primary limitations of our current model. 
First, our model assumes the stochastic forcing acts as Gaussian white noise. While this successfully models average, gradual orbital diffusion, it overlooks the rare but extreme perturbations present in realistic disks. As \citet{Laughlin2004} demonstrated, large-scale fluid structures can occasionally exert massive, sudden torques that rapidly disrupt resonances, whereas a purely Gaussian distribution treats these powerful events as almost impossible. Therefore, while our framework provides a solid baseline, realistic disk turbulence may disrupt resonances more efficienty. Incorporating these infrequent but extreme events represents an important direction for our future work.

Second, we treat the strength of the stochastic force $\kappa$ and the timescales of angular momentum and eccentricity damping $\tau_{\rm m}$ and $\tau_{\rm e}$ as independent parameters. However, in a realistic protoplanetary disk, these quantities are intrinsically coupled, as they share dependencies on underlying disk properties such as the gas surface density, aspect ratio, and local turbulent viscosity. A more self-consistent approach would require linking these parameters explicitly within a unified disk model.

Besides, our simulations assumed a static background disk. In reality, protoplanetary disks evolve over several Myr timescales, experiencing significant gas depletion. Accounting for such time-dependent disk evolution represents a crucial next step, particularly for connecting our dynamical predictions to the observed statistical properties of exoplanet populations.

\section{Conclusion}
\label{sec:conclusion}
In this study, we extend the theoretical framework of Paper I \citep{Lin2025} to turbulent disks, establishing an analytical model for the capture and stability of two-planet systems undergoing convergent migration. Planet–disk interactions are parameterized into two components: smooth migration in a classical laminar disk, characterized by angular-momentum and eccentricity-damping timescales ($\tau_{\rm m}$, $\tau_e$), and stochastic migration arising from turbulent fluctuations in the disk potential, quantified by the forcing strength $\kappa$ (Eq. \ref{eq:kap_def}). Using this formulation, we derive an analytical stability criterion (Eq. \ref{eq:final_kappa}) for general $j$:$j-1$ resonances that is valid for arbitrary planet mass ratios.

The dynamical outcomes can be classified into two distinct regimes (Figure \ref{fig:illustration}): (i) resonant trapping; and (ii) resonance disruption induced by turbulence. These regimes and their boundaries, validated through extensive $N$-body simulations, are delineated in the $\kappa$-$\tau_{\rm m}/\tau_{e}$ parameter space (Figures \ref{fig:kap_crit}, \ref{fig:kap_crit_q_1} and \ref{fig:kap_crit_q_10}).

Our analysis reveals that turbulence systematically reduces the stability of resonant systems. In the laminar-disk limit (Paper I), resonance escape via overstability occurs when the migration-to-damping timescale ratio $\tau_{\rm m}/\tau_{e}$ falls below a critical value, driven by excessive eccentricity excitation. We demonstrate that turbulent diffusion acts as an additional source of excitation that exacerbates resonant overstability. Consequently, to counteract this stochastic forcing, the system necessitates stronger eccentricity damping (i.e., a higher $\tau_{\rm m}/\tau_{e}$) than predicted by laminar theory. Specifically, increasing turbulence effectively expands the parameter space for overstability, shifting the critical escape threshold toward higher $\tau_{\rm m}/\tau_{e}$ values. Ultimately, when diffusive perturbations are sufficient to exceed the resonance width, stable capture is precluded regardless of the damping rate.

\begin{acknowledgements}
The authors appreciate the constructive comments from the referee.  BL is supported by the National Key R\&D Program of China (2024YFA1611803), the National Natural Science Foundation of China (Nos. 12222303 and 12173035), and the start-up grant of the Bairen program from Zhejiang University. M.H.L. was supported in part by the Hong Kong RGC grant 17309323. HY is supported by the National
Natural Science Foundation of China (Grant No. 12473067). The simulations and analysis presented in this article were carried out on the SilkRiver Supercomputer of Zhejiang University.
Software: \texttt{REBOUND} \citep{rein2012} and \texttt{REBOUNDX} \citep{Tamayo2020}.
\end{acknowledgements}

\bibliographystyle{aa}
\bibliography{main.bib}

\begin{appendix}

\section*{Appendix A: Derivation of the resonance width}
\label{sec:appendix A}
\renewcommand{\theequation}{A\arabic{equation}}
\setcounter{equation}{0}

For a first-order $j:j-1$ mean motion resonance, the resonant angle is defined as $\varphi = j\lambda_{\rm o} - (j-1)\lambda_{\rm i} - \varpi_{\rm i,o}$, where $\lambda$ is the mean longitudes of the planet, and $\varpi$ is the relevant longitude of pericenter. the resonant dynamics can be simplified to a classical pendulum model:
\begin{equation}
    \ddot{\varphi} + \omega_0^2 \sin \varphi = 0,
\end{equation}
where $\omega_0$ is the libration frequency and $\tau_{\rm lib} = 2\pi/\omega_0$ is the associated period. 

Under convergent Type I migration, the evolution of the mean motion ratio $x = n_i/n_o - j/(j-1)$ is governed by the migration timescale $\tau_{\rm m}$. The rate of change of the ratio near the resonance is given by:
\begin{equation}
    \dot{x} = - \frac{3}{\tau_{\rm m}} \left( \frac{n_{\rm i}}{n_{\rm o}} \right).
\end{equation}

The resonance width $x_{\rm res}$ represents the range of the mean motion ratio over which capture is probable. In the adiabatic limit, this width corresponds to the change in $x$ accumulated over a single libration period at the critical migration limit $\tau_{\rm slow,mig}$:
\begin{equation}
    x_{\rm res} = |\dot{x}| \tau_{\rm lib} = \frac{3n_{\rm i}}{n_{\rm o}} \frac{\tau_{\rm lib}}{\tau_{\rm slow,mig}}.
\end{equation}
where \citep{Lin2025}
\begin{equation}
 \begin{cases}
 {\displaystyle   
 \tau_{\rm lib} = \frac{2\pi}{\sqrt{3} [(j-1)^{2}\mu_{\rm o}n_{\rm i}^{2}\alpha + j^{2}\mu_{\rm i}n_{\rm o}^{2}]^{1/3}} \cdot \frac{1}{(\mu_{\rm o}n_{\rm i}\alpha f_{\rm d,i}^{2} + \mu_{\rm i}n_{\rm o}f_{\rm d,o}^{'2})^{1/3}},   } \\
   {\displaystyle    }  
\vspace{0.1cm}\\
 {\displaystyle       \tau_{\rm slow,mig} = \frac{1}{[ (j-1) \mu_{\rm o} n_{\rm i} \alpha + j \mu_{\rm i} n_{\rm o} ]} \frac{[(j-1)^{2}\mu_{\rm o}n_{\rm i}^{2}\alpha + j^{2}\mu_{\rm i}n_{\rm o}^{2}]^{1/3}}{3^{1/3}(\mu_{\rm o}n_{\rm i}\alpha f_{\rm d,i}^{2} + \mu_{\rm i}n_{\rm o}f_{\rm d,o}^{'2})^{2/3}}.  }\\
     {\displaystyle   } 
\end{cases}
\label{eq:trap_criteria}
\end{equation}
This analytical expression for $x_{\rm res}$ serves as the theoretical basis for the resonance disruption criterion in Section \ref{sec:tur_crit}.

\section*{Appendix B: Effect of eccentricity damping on resonance survival}
\label{app:appendix_B}

\renewcommand{\theequation}{B\arabic{equation}}
\setcounter{equation}{0}

\renewcommand{\thefigure}{B.\arabic{figure}}
\setcounter{figure}{0}

In the main text, we estimate the stochastic diffusion within resonance using \(\sigma_x^2 \simeq D_x\tau_{\rm lib}\). This approximation treats the variance growth as locally linear over the resonant response time, but does not explicitly include the filtering effect of eccentricity damping in the effective diffusion coefficient. Here we test the possible influence of \(\tau_e\) on resonance survival with an ensemble of numerical experiments.

We focus on the small-outer-planet case with \(m_{\rm i}=10\,M_{\oplus}\), \(m_{\rm o}=1\,\ M_{\oplus}\),\ \(q=10\), and \(\tau_{\rm m}=8\times10^{5}\,{\rm yr}\). For this setup, the libration timescale estimated from
Eq.~\ref{eq:trap_criteria} is \(\tau_{\rm lib}\simeq10^3\,{\rm yr}\). We vary \(\tau_{\rm m}/\tau_e=5\times10^2, 10^3, 1.5\times10^3\), and \(2\times10^3\). For each value, we perform 20 simulations and measure the fraction of systems that remain near the 2:1 resonance as a function of time. Resonance survival is identified using $\left|P_{\rm o}/P_{\rm i}-2\right|<0.05$.

\begin{figure}
    \centering
    \includegraphics[width=0.55\textwidth]{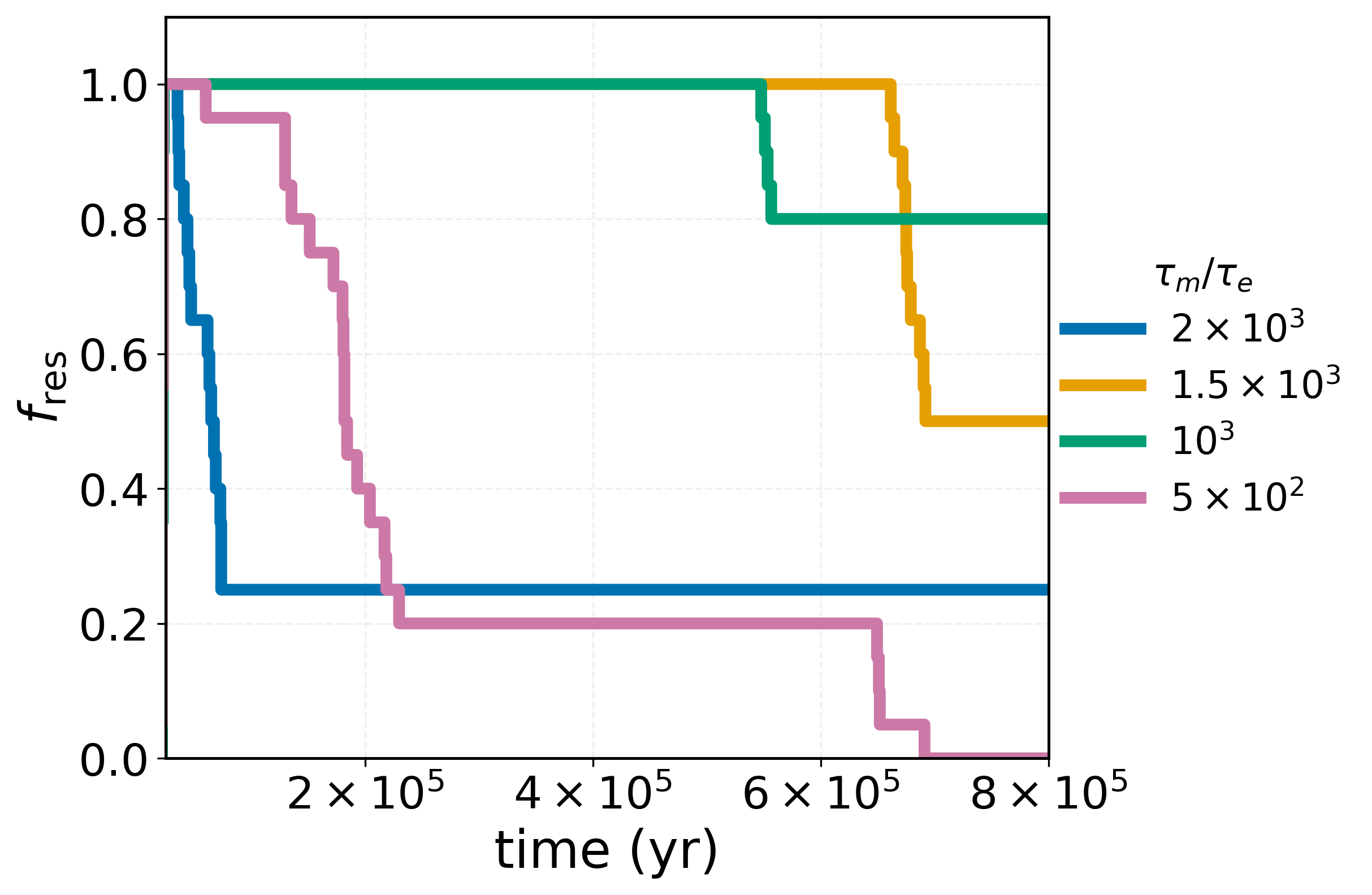}
    \caption{
    Fraction of systems that remain near the 2:1 resonance as a function of time for different values of \(\tau_{\rm m}/\tau_e\). The simulations adopt \(m_{\rm i}=10\,M_{\oplus}\), \(m_{\rm o}=1\,M_{\oplus}\), \(q=10\), and \(\tau_{\rm m}=8\times10^{5}\,{\rm yr}\). Each curve is computed from 20 stochastic realizations. Resonance survival is identified using \(|P_{\rm o}/P_{\rm i}-2|<0.05\).
    }
    \label{fig:res_frac}
\end{figure}

Figure~\ref{fig:res_frac} shows the time evolution of the resonance survival fraction for different values of \(\tau_{\rm m}/\tau_e\). For \(\tau_{\rm m}/\tau_e=5\times10^2\), most systems eventually leave the resonance, whereas for \(\tau_{\rm m}/\tau_e=10^3\), most systems remain resonant throughout the integration. The \(\tau_{\rm m}/\tau_e=1.5\times10^3\) and \(2\times10^3\) cases show partial disruption with different disruption times. The trend is not strictly monotonic, likely because of the limited number of stochastic realizations and the complex role of eccentricity damping in the resonant dynamics. 
\\
Since the leading-order analytical formula already provides a good match to the numerical results, we retain the analytical approximation of  \(\sigma_x^2\simeq D_x\tau_{\rm lib}\) in the main text without introducing additional empirical corrections.

\end{appendix}

\end{document}